# Chiral phonons probed by X rays

Hiroki Ueda[1,2,*], Mirian García-Fernández[3], Stefano Agrestini[3], Carl P. Romao[4], Jeroen van den Brink[5,6], Nicola A. Spaldin[4], Ke-Jin Zhou[3], and Urs Staub[1,*]

[1] *Swiss Light Source, Paul Scherrer Institute, 5232 Villigen-PSI, Switzerland.*

[2] *SwissFEL, Paul Scherrer Institute, 5232 Villigen-PSI, Switzerland.*

[3] *Diamond Light Source, Harwell Campus, Didcot OX11 0DE, United Kingdom.*

[4] *Department of Materials, ETH Zurich, 8093 Zürich, Switzerland.*

[5] *Institute for Theoretical Solid State Physics, IFW Dresden, Helmholzstr. 20, 01069 Dresden, Germany.*

[6] *Institute for Theoretical Physics and Würzburg-Dresden Cluster of Excellence ct.qmat, Technische Universität Dresden, 01069 Dresden, Germany.*

[*] Correspondence authors: hiroki.ueda@psi.ch and urs.staub@psi.ch

**The concept of chirality is of great relevance in nature, from chiral molecules such as sugar to parity transformations in particle physics. In condensed matter physics, recent studies have demonstrated chiral fermions and their relevance in emergent phenomena closely related to topology [1-3]. The experimental verification of chiral phonons (bosons) remains challenging, however, despite their expected strong impact on fundamental physical properties [4-6]. Here we show experimental proof of chiral phonons using resonant inelastic X-ray scattering with circularly polarized X rays. Using the prototypical chiral material, quartz, we demonstrate that circularly polarized X rays, which are intrinsically chiral, couple to chiral phonons at specific positions in reciprocal space, allowing us to determine the chiral dispersion of the lattice modes. Our experimental proof of chiral phonons demonstrates a new degree of freedom in condensed matter that is both of fundamental importance and opens the door to exploration of novel emergent phenomena based on chiral bosons.**

Quasiparticles in solids fundamentally govern many physical properties, and their symmetry is of central importance. Chiral quasiparticles are of particular interest. For example, chiral fermions emerge at degenerate nodes in Weyl semimetals [1] and chiral crystals [2,3]. Their chiral characters are directly manifested by a chiral anomaly [7] and lead to enriched topological properties, including selective photoexcitation by circularly polarized light [8], chiral photocurrent [9], and transport [7]. The presence of chiral bosons such as phonons [4-6,10-17] and magnons [6,18-20], has also extensively been debated.

Chiral phonons are vibrational modes of solids in which the atoms have a rotational motion perpendicular to their propagation with an associated circular polarization and angular momentum. As a result of their angular momentum, chiral phonons can carry orbital magnetic moments, enabling a phono-magnetic effect analogous to the opto-magnetic effect from other helical atomic rotations [21,22]. Correspondingly, the phonons can create an effective magnetic field, which has been invoked to explain the observation of excited magnons [23] and enables their excitation through ultrafast angular-momentum transfer from a spin system [24]. Whereas a phononic magnetic field has so far been discussed primarily at the $\Gamma$ point, chiral phonons naturally arise in noncentrosymmetric materials away from the zone center, and are based on a fundamentally different symmetry.

Experimental observation of phonon chirality has proven to be challenging. If atomic rotations are confined in a plane containing the phonon propagation direction (circular phonons), the mode cannot possess a chiral character (see Supplementary Information for symmetry consideration), as occurs for non-propagating phonons at $\Gamma$ and other high-symmetry points. Therefore, results based on optical-probe techniques, such as chiroptical spectroscopy [16] and circularly polarized Raman scattering [17], are insufficient to identify the presence of chiral phonons because of the large wavelength of optical photons, restricting the exploration very close to the $\Gamma$ point. The first claim of observation of a chiral phonon was made at the high-symmetry points of a monolayer transition-metal dichalcogenide [5], though it has been argued to be inconsistent with symmetry arguments [6]. Thus, establishing an experimental method that directly verifies the chiral character of phonons is strongly demanded.

In this work, we demonstrate chiral phonons in a chiral material at general momentum points in the Brillouin zone. We probe the chirality of phonons using resonant inelastic X-ray scattering (RIXS) with circularly polarized X rays. Our strategy rests on the fact that circularly polarized X rays are chiral and is inspired by the use of resonant *elastic* X-ray scattering to probe the chirality of a *static* lattice by using circularly polarized X rays on screw-axis forbidden reflections [25]. Using RIXS, circularly polarized chiral photons can couple to *dynamic* chiral phonon modes by transferring angular momentum, and the process can occur at general momentum points in reciprocal space. Our theoretical analysis shows that the observed circular dichroism in RIXS is caused by the orbitals of the resonant atoms that align in a chiral way determined by the chiral crystal structure; we calculate the angular momentum of the phonons at the corresponding **Q** point using density-functional theory (DFT).

RIXS is a two-step process in which the energy of the incident photon with a given polarization coincides (resonates) with an atomic X-ray absorption edge of the system [26]. For RIXS at the O $K$ edge, an incident photon excites an electron from the O $1s$ inner shell to the $2p$ outer shell. The combined core hole and excited electron form a short-lived excitation in this intermediate state that interacts with the lattice and creates phonons as it deforms its local environment [27,28]. The final RIXS step involves the de-excitation of the electron from $2p$ to $1s$, causing the emission of a photon while leaving behind a certain number of phonons in the system. The detected energy and momentum of the emitted photon are directly related to the energy and momentum of the phonon created in the solid.

To illustrate the mechanism by which RIXS excites chiral phonons in quartz, we consider a Si-O chain in which the O ions bond to the Si ions via the $2p$ orbital pointing towards the central axis of the chain (see Fig. 1 and Fig. S1). While this O 2p orbital is unchanged in the local frame of the ligand as it rotates around the central axis with angle $\phi$, in the global frame its direction changes upon rotation. We describe the spatial coordinate of the phonon by the angle $\phi$ and denote the creation operator of an electron in the 2p orbital along the global x axis as $p_x^\dagger$ and along the global y axis as $p_y^\dagger$. We construct the RIXS intermediate state Hamiltonian $H_I$ such that during an (adiabatically slow) rotation of the atom around the z axis, the ground state wavefunction always points towards the center of rotation (see Supplementary Information for a detailed derivation):

$$H_I = -\alpha s s^\dagger \mathbf{p}^\dagger (\sigma_z \cos 2\phi + \sigma_x \sin 2\phi) \mathbf{p}, \quad (1)$$

where $ss^\dagger$ is the core-hole density operator, the vector operator $\mathbf{p} = (p_x, p_y)$ and $\sigma_i$ denotes the Pauli matrixes with $i = x, y, z$. The RIXS operator that takes the system from the ground state $|0>$ to the final state $|f>$ with $m$ phonon modes can be evaluated to lowest order in $\alpha$ using the ultrashort core-hole lifetime expansion [27]. Introducing the circular polarization basis $\boldsymbol{\epsilon}_c$, where a fully left circularly polarized photon corresponds to $\boldsymbol{\epsilon}_c^L = (1,0)$ and a right one to $\boldsymbol{\epsilon}_c^R = (0,1)$, the RIXS amplitude becomes (see Supplementary Information)

$$A_m = (\boldsymbol{\epsilon}_c')^* \langle m | \sigma_z e^{2i\phi\sigma_y} |0\rangle \boldsymbol{\epsilon}_c. \quad (2)$$

This shows that angular momentum is transferred to the phononic system when the incident and scattered photons have different circular polarization. Figure 1 shows conceptually how such interactions between circularly polarized photons and the lattice can launch rotational lattice vibrations through this angular momentum transfer.

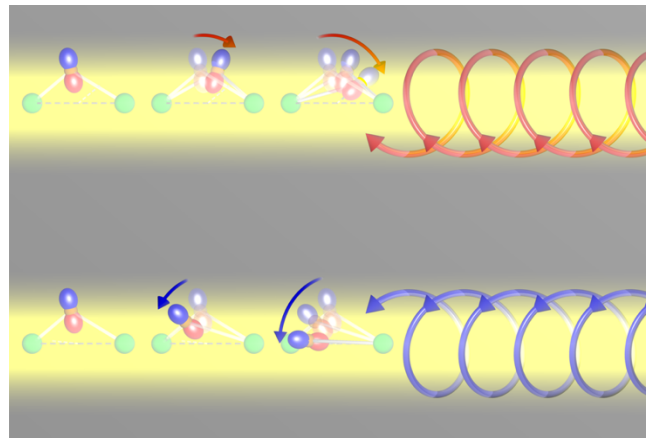

**Fig. 1 | Angular-momentum transfer in a RIXS experiment.** The angular momentum of the photons [opposite between C+ (up, red) and C- (down, blue)] is transferred to a crystal, causing a rotation in this case of anions (orange spheres with $p$ orbitals) relative to their neighboring cations (green spheres).

As our target material, we choose the prototypical chiral crystal, quartz (α-SiO₂), in which SiO₄ tetrahedra form a chiral helix along [001] (Fig. 2). The resulting chiral space group is either $P3_221$ (left quartz, Fig. 2a) or $P3_121$ (right quartz, Fig. 2b). A recent DFT study [15] pointed out the chirality and phonon angular momentum of some phonon branches and demonstrated the reversal of chirality between opposite enantiomers, as well as the absence of phonon angular momentum at the $\Gamma$ point.

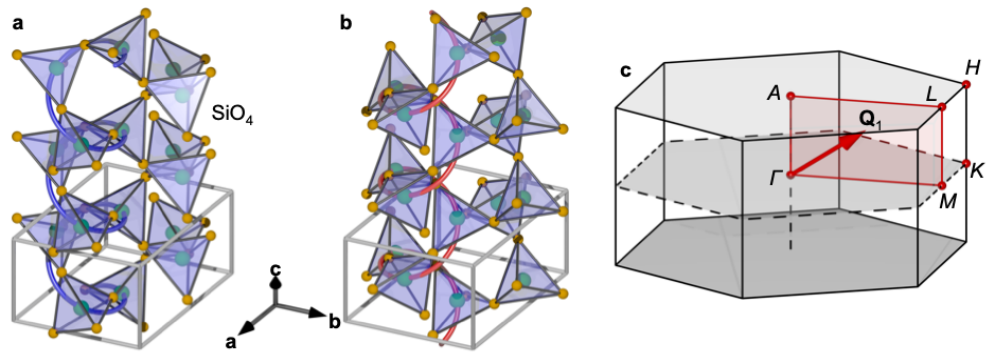

**Fig. 2 | Crystal structure and Brillouin zone of quartz.** Crystal structures of **a**, left quartz and **b**, right quartz, and **c**, the Brillouin zone with $\mathbf{Q}_1$, where the RIXS spectra has been taken.

We performed RIXS experiments with circular polarization (C+/C-) on two quartz crystals with opposite chirality. With incident photon energy tuned around the O $K$ edge reaching an energy resolution of ~28 meV, we collected spectra at $\mathbf{Q}_1$ = (-0.25, 0, 0.32) r.l.u. (see Fig. 2c) (see Methods for details). The spectrum for various incident photon energies (shown in Fig. 3) shows clear peaks on the energy-loss side at resonance, which become suppressed for energies further away from resonance. Note that the energy resolution is insufficient to assign the peaks to individual phonons [29]. All peaks above the energy of the highest phonon mode of ~0.2 eV [29] are the result of higher-harmonic phonon excitations.

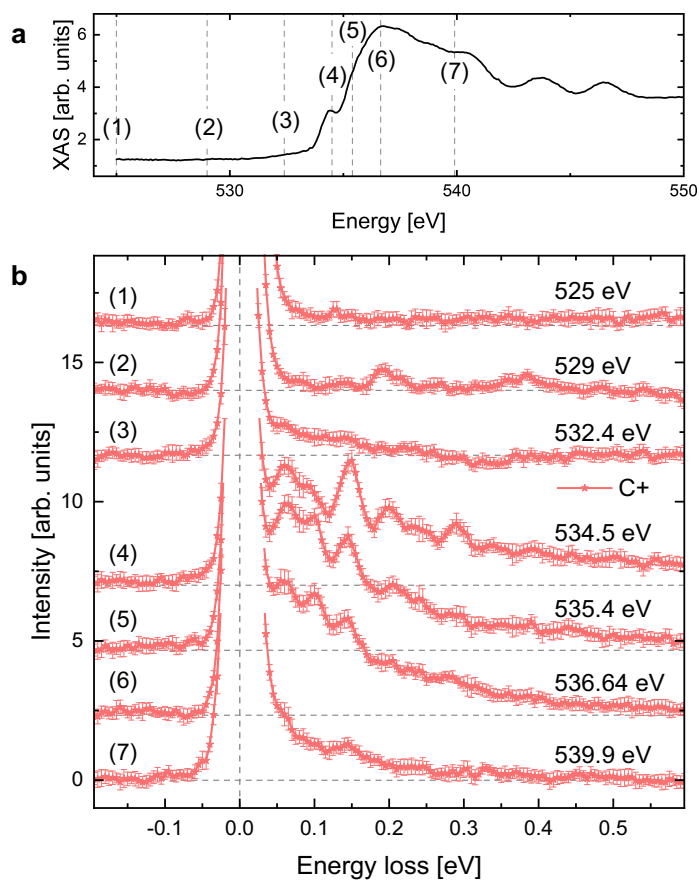

**Fig. 3 | XAS and photon-energy dependence of RIXS. a**, X-ray absorption spectrum around the O $K$ edge, and **b**, RIXS spectra taken with C+ for left-handed quartz at $\mathbf{Q}_1$ = (-0.25, 0, 0.32) for the incident photon energies indicated by the dashed lines in **a**. Each spectrum in **b** is vertically shifted to enhance visibility. Errorbars are smaller as the line (a) and in Standard Deviation (SD) (b).

Figure 4 shows the C+ and C- RIXS spectra from left- (4a) and right- (4b) handed quartz and their dichroic contrasts (4c) at 20 K. We see a clear contrast between C+/C-, and the dichroism changes sign for the opposite chiral enantiomers, indicating that it is caused by chirality of the modes. We find similar contrast between C+/C- at the other reciprocal points with different RIXS spectra due to different phonon energies (dispersion) and different RIXS cross sections. (see Fig. S2 in Supplementary Information). These observations demonstrate unambiguously that circularly polarized photons couple to chiral phonons, with the chirality of the phonons defined by the lattice chirality, and that RIXS with circularly polarized X rays can be used to probe phonon chirality.

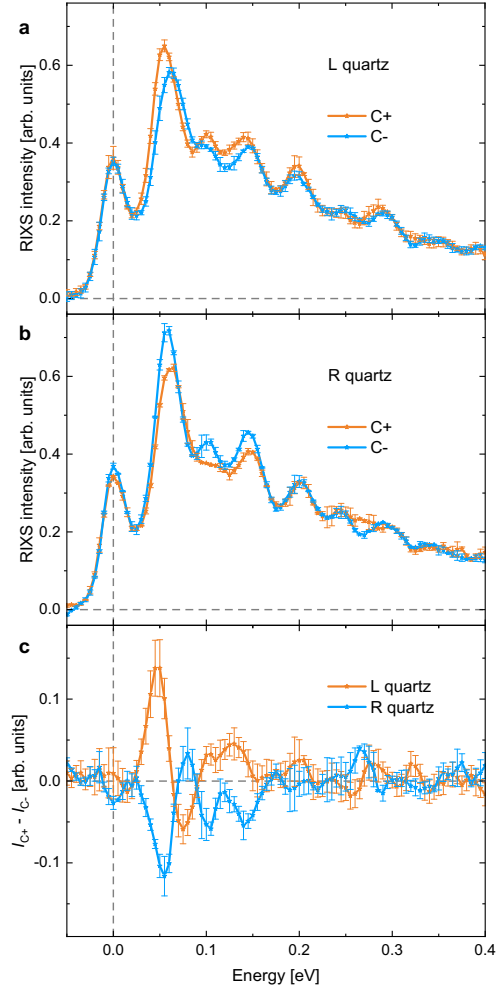

**Fig. 4 | RIXS with circularly polarized X rays.** Comparison between **a**, left quartz and **b**, right quartz, taken at the incident photon energy of 534 eV and $Q_1$ = (-0.25, 0, 0.32). **c**, Extracted circular dichroic components of the data shown in **a** and **b**. Errorbars in SD

We use DFT to calculate the phonon dispersion and phonon circular polarization for all phonon branches, and show their dispersion between $Q_1$ and $\Gamma$ in Fig. 5a for left quartz (details found in Methods, see Figs. S4 and S5 for other directions in reciprocal spaces and components of the circular polarization vector). Note that, since we are interested in low symmetry points in the Brillouin zone, we show a different direction from that in Ref. [15], as well as additional bands. The color scale indicates the phonon circular polarization (**S**) [4], which indicates the chirality of a phonon mode; it is defined, for example for the *z* component $S_z$, as

$$S_z = \sum_{m=1}^{n} S_{z,m} = \sum_{m=1}^{n}\left(|\langle r_{m,z}|\epsilon_m\rangle|^2 - |\langle \ell_{m,z}|\epsilon_m\rangle|^2\right). \quad (3)$$

Here $\epsilon_m$ are the phonon eigenvectors of each atom (normalized such that $\sum_m |\langle\epsilon_m|\epsilon_m\rangle| = 1$), and $|r_{m,z}\rangle$ and $|\ell_{m,z}\rangle$ are eigenvectors corresponding to pure right- and left-handed rotations. The phonon angular momentum (**L**) is then given by $\mathbf{L} = \hbar\mathbf{S}$ [4]. We also report the mode

effective charges (Fig. 5b) as a metric of the strength of the interaction between the mode and light, calculated following the method of Ref. [30].

When we match the calculated and measured modes, we find that those with the strong dichroic contrast are those calculated to have a large chirality. The peak with the largest contrast is at ~50 meV for all the reciprocal points we measured [$Q_1$ in Fig. 4c, and $Q_2$ = (-0.29, 0.14, 0.32) and $Q_3$ = (-0.25, 0.25, 0.32) shown in Fig. S2], suggesting that a mode that has large phonon circular polarization and energy around 50 meV dominates the contrast. The mode at the energy of ~47.6 meV at $Q_1$, which we refer to as mode X, matches the conditions (see Supplementary Figure 4 and Supplementary Table 1, which tabulates the energy and phonon circular polarization of all phonon modes at the measured $Q$ points). Figure 5c and Supplementary Movie 1 visualize mode X at $Q_1$, and show that it involves a circular motion of the atoms. Importantly, the mode satisfies the symmetry requirement for a chiral phonon mode.

For non-magnetic quartz, the RIXS spectra at the O $K$ edge are mainly sensitive to the O $2p$ orbital states. This means that phonon modes that significantly affect for example the orientation of the $2p$ orbital states will create large scattering contrast in RIXS, and will also be strongly X-ray polarization dependent. Figure 5d and Supplementary Movie 2 visualize the evolution of the local charge quadrupoles at the O site when the chiral phonon mode is excited (see Fig. 5c or Supplementary Movie 1). These charge quadrupoles reflect the time evolution of the O $2p$ orbitals, which shows that the dichroic RIXS signal is due to an evolution of the chiral stacking of the O $2p$ orbital moments in the chiral phonon excitation as described in (1) and (2) above.

Note that mode with the largest contrast is not the phonon mode with the largest phonon circular polarization. Instead, the mode has a large mode effective charge at $Q_1$, as shown in Fig. 5b. This indicates that the contrast depends on not only the chiral amplitude of a mode itself but also the modulation of the electronic charges with respect to the plane of the electric fields of the circularly polarized X rays. Note that there is an additional consideration: Phonon circular polarization specifies a preferred rotation direction of atoms in the excitation, which can only be excited with the matching circular photon polarization (see Fig. 1). As modes of opposite chirality have different energies (see Fig. S4 and Supplementary Table 1 that modes with opposite chirality, degenerated at the $\Gamma$ point, split at away from the zone center), the peaks which are composed of several modes show a peak shift when taken with opposite circular polarization (see Fig. 4).

In Fig. 5e we show the associated magnetic moments induced by the chiral motion of the charged ions in the chiral phonons, which we calculate by extending the method used in Refs. [21,22] so that it is applicable at an arbitrary point in $Q$-space. We begin by constructing the atomic circular polarization vector $S_m$ as $S_m = [S_{x,m}\ S_{y,m}\ S_{z,m}]$ [see (3)]; yielding the angular momentum of each atom as $L_m = \hbar S_m$. The magnetic moment ($\mu_m$) of each atom participating in the phonon is

$$\mu_m = L_m \gamma_m = \hbar S_m Z_m / 2 m_m, \quad (4)$$

where $\boldsymbol{\gamma}_m$ is the gyromagnetic ratio tensor, which is derived from $\boldsymbol{Z}_m$, the Born effective charge tensor, and $m_m$, the atomic masses. The phonon magnetic moment is then simply $\boldsymbol{\mu} = \sum_{m=1}^{n} \boldsymbol{\mu}_m$. We show our calculated mode- and **Q**-point resolved magnetic moments in Fig. 5e and see that chiral phonons in quartz carry magnetic moments throughout the Brillouin Zone, although the calculated magnetic moments are relatively small due to the low values of $\boldsymbol{\gamma}_m$. These phonon magnetic moments do not normally create a net magnetization due to the presence of time-reversal related pairs with opposite chirality and magnetic moment. If time-reversal symmetry is broken, however, population imbalances between the chiral pairs can be created [31]. Figure 5e also suggests that the phonon chirality can be investigated directly through interactions with the phonon magnetic moment, using, e.g., polarized inelastic neutron scattering.

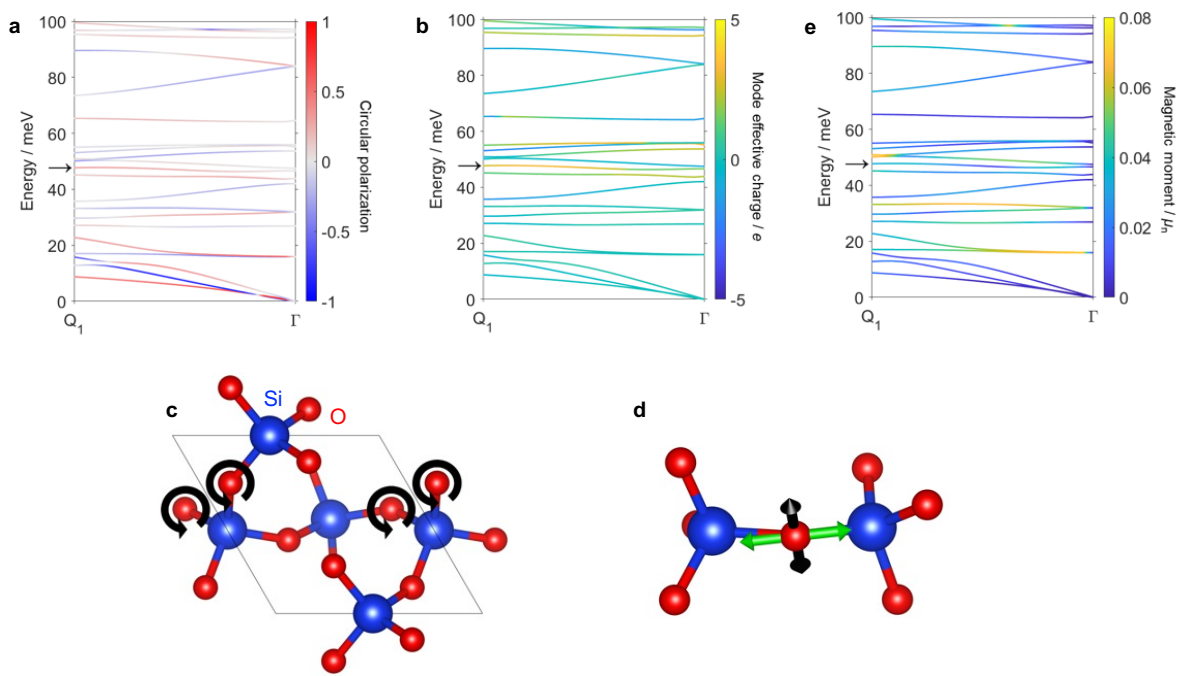

**Fig. 5 | Phonon dispersion and chiral phonon mode. a**, Low-energy phonon dispersion for left quartz along the $\Gamma - Q_1$ direction. Colors represent the $z$ component of the phonon circular polarization. **b**, The same phonon band structure with colors representing mode effective charges (a measure of the degree in which the electronic charge distribution is perturbed by the phonons), in units of the elementary charge. **c**, The chiral phonon mode at $\mathbf{Q}_1 = (-0.25, 0, 0.32)$ (indicated with an arrow in **a**) showing the main chiral revolutions of the oxygen atoms that have a different phase along the chain. **d**, Associated change in the local quadrupole moment (associated with the O 2p orbital) for a revolving oxygen atom between the phonon at phase 0 and phase $\pi$ (black vectors representing an increase in the atomic quadrupole moment between its position at phonon phase 0 and its position at phase $\pi$ and green vectors representing a decrease). **e**, The phonon band structure colored according to the magnitude of the magnetic moment of the phonons, in units of the nuclear magneton.

In conclusion, we have used resonant inelastic X-ray scattering with circularly polarized X rays to demonstrate the chiral nature of the phonons in chiral quartz crystals, and in turn have established a fundamental methodology for characterizing chiral phonons. With the technique established by this proof-of-principle study, the chirality of phonons at general momentum points can be characterized, opening up new perspectives in chiral phononics. For example, our work indicates that RIXS can be used to quantify the role of chiral phonons in exotic phenomena proposed in topological materials [32-35], as well as to characterize interactions such as electron- and spin- couplings with chiral phonons [14,36-39].

## Methods

**Resonant inelastic X-ray scattering.** RIXS measurements were performed at the Beamline I21 of Diamond Light Source in the UK [41]. Used photon energy is around the O $K$ edge, and polarization is circular (C+/C-). The energy resolution is estimated as 28 meV from the full width of the half-maximum of the elastic peak from a carbon tape. Enantiopure single crystals purchased commercially, have the widest face perpendicular to the [001] axis. The manipulator installed at the beamline allows us to rotate the crystal along the azimuthal angle, enabling us to access different momentum points while the experiment, $Q_1$ = (-0.25, 0, 0.32), $Q_2$ = (-0.29, 0.14, 0.32), and $Q_3$ = (-0.25, 0.25, 0.32). XAS obtained prior to the RIXS measurements is based on the total electron yield method.

**Density-functional theory.** Density-functional calculations were performed using the Abinit software package (v. 9) [42,43] and the Perdew–Burke–Ernzerhof exchange–correlation functional [44] with the dispersion correction of Grimme [45]. The phonon band structure was determined using density functional perturbation theory [42], using norm-conserving pseudopotentials, a 38 Ha plane wave energy cutoff, an 8 × 8 × 8 Monkhorst–Pack grid in **k**-space [46], and a 4 × 4 × 4 grid in **Q**-space. Calculations of the electronic and phononic structure were additionally performed explicitly at the experimentally measured **Q** points. Frozen-phonon calculations were performed using the projector-augmented wave (PAW) method [47] to obtain local quadrupole moments with the multipyles post-processing script [48]. These calculations used a 192 Ha plane wave energy cutoff within the atomic spheres and a 32 Ha cutoff without. The default pseudopotentials and PAW datasets from the Abinit library were used.

## Data availability

Experimental and model data are accessible from the PSI Public Data Repository [40].

**Acknowledgments**

We would like to thank A. Nag for advising on the data analysis and stimulating discussion. The resonant inelastic X-ray scattering experiments were performed at the I21 beamline in the Diamond Light Source under proposal No. MM28375. H.U. was supported by the National Centers of Competence in Research in Molecular Ultrafast Science and Technology (NCCR MUST-No. 51NF40-183615) from the Swiss National Science Foundation and from the European Union's Horizon 2020 research and innovation program under the Marie Skłodowska-Curie Grant Agreement No. 801459 – FP-RESOMUS. This work was funded by the European Research Council under the European Union's Horizon 2020 Research and Innovation Program, Grant Agreement No. 810451. Computational resources were provided by ETH Zurich and the Swiss National Supercomputing Center (CSCS) under project ID eth3. C.P.R. acknowledges the support of the European Union and Horizon 2020 through the Marie Skłodowska-Curie Fellowship, Grant Agreement No. 101030352. J.v.d.B. thanks the Deutsche Forschungsgemeinschaft (DFG) for support through the Würzburg-Dresden Cluster of Excellence on Complexity and Topology in Quantum Matter ct.qmat (EXC 2147, Project No. 39085490) and the Collaborative Research Center SFB 1143 (Project No. 247310070).

**Author contributions**

H.U. and U.S. conceived and designed the project. H.U., M.G.F., S.A., K-J.Z., and U.S. performed resonant inelastic X-ray scattering experiments. H.U. analyzed the experimental data. C.P.R. and N.A.S. performed density functional theory calculations. J.v.d.B. contributed to the mechanism by which RIXS excites chiral phonons. H.U., C.P.R., J.v.d.B., N.A.S., and U.S. wrote the manuscript with contributions from all authors.

**Additional information**

**ORCIDs**

| | |
|---|---|
| Hiroki Ueda: | 0000-0001-5305-7930 |
| Mirian Garcia-Fernandez: | 0000-0002-6982-9066 |
| Stefano Agrestini: | 0000-0002-3625-880X |
| Carl Peter Romao: | 0000-0002-6233-8133 |
| Jeroen van den Brink: | 0000-0001-6594-9610 |
| Nicola Spaldin: | 0000-0003-0709-9499 |
| Kejin Zhou: | 0000-0001-9293-0595 |
| Urs Staub: | 0000-0003-2035-3367 |